# Causal Inference for Survival Analysis

Vikas Ramachandra

Stanford University Graduate School of Business

655 Knight Way, Stanford, CA 94305

## Abstract

In this paper, we propose the use of causal inference techniques for survival function estimation and prediction for subgroups of the data, upto individual units. Tree ensemble methods, specifically random forests were modified for this purpose. A real world healthcare dataset was used with about 1800 patients with breast cancer, which has multiple patient covariates as well as disease free survival days (DFS) and a death event binary indicator (y). We use the type of cancer curative intervention as the treatment variable (T=0 or 1, binary treatment case in our example).

The algorithm is a 2 step approach. In step 1, we estimate heterogeneous treatment effects using a causalTree with the DFS as the dependent variable. Next, in step 2, for each selected leaf of the causalTree with distinctly different average treatment effect (with respect to survival), we fit a survival forest to all the patients in that leaf, one forest each for treatment T=0 as well as T=1 to get estimated patient level survival curves for each treatment (more generally, any model can be used at this step).

Then, we subtract the patient level survival curves to get the differential survival curve for a given patient, to compare the survival function as a result of the 2 treatments.

The path to a selected leaf also gives us the combination of patient features and their values which are causally important for the treatment effect difference at the leaf.

## 1. The problem of causal inference

We consider a setup where there are n units or data points, indexed by i = (1, . . . , n). We postulate the existence of a pair of potential outcomes for each unit, $(Y_i(0), Y_i(1))$ (following the potential outcome or Rubin Causal Model [4]), with the unit-level causal effect defined as the difference in potential outcomes, $T_i = Y_i(1) - Y_i(0)$. Let $W_i \in \{0, 1\}$ be the binary indicator for the treatment, with $W_i = 0$ indicating that unit i received the control treatment, and $W_i = 1$ indicating that unit i received the active treatment. The realized outcome for unit i is the potential outcome corresponding to the treatment received: $Y(obs)_i = Y_i(W_i) = Y_i(0) \text{ if } W_i = 0, Y_i(1) \text{ if } W_i = 1$.

Let $X_i$ be a N-component vector of features, covariates or pretreatment variables, known not to be affected by the treatment. Our data consist of the triple $(Y(obs)_i, W_i, X_i)$, for i = (1, . . . , n), which are regarded as an i.i.d sample drawn from a large population. We assume that observations are exchangeable, and that there is no interference (the stable unit treatment value assumption, or sutva).

Since we cannot observe the counterfactual for any particular $x_i$ unit, one way to estimate the treatment effect for each unit will be by using values from its neighbors which received the opposite treatment, and by taking the difference between the two outcomes. This individual treatment effect ITE can be written as:

$ITE =$
$T(estimated)_i = Y_i(1) - Y_{neighbor}(0)$, if $W_i = 1$, and $-(Y_i(0) - Y_{neighbor}(1))$, if $W_i = 0$

There are different techniques to determine the 'neighbors' in the above construct, and in this paper, we use the causalTree approach form clusters of 'neighbors' in the form of leaves, and estimate the treatment effect at each leaf [5]. In the next section, we look at survival models.

## 2. Survival analysis and data heterogeneity

Survival analysis is a set of statistical methods designed for analyzing the expected duration of time until one or more events happen, such as death in humans.
The following terms are commonly used for survival analysis.
Event: Death, disease occurrence, disease recurrence, recovery, or other experience of interest
Time: The time from the beginning of an observation period (such as surgery or beginning treatment) to (i) an event, or (ii) end of the study, or (iii) loss of contact or withdrawal from the study.
Censoring / Censored observation: If a subject does not have an event during the observation time, they are described as censored. The subject is censored in the sense that nothing is observed or known about that subject after the time of censoring. A censored subject may or may not have an event after the end of observation time.
Survival function S(t): The probability that a subject survives longer than time t.
The survival function can be estimated using either parametric (such as an exponential function) or non parametric techniques.
The Kaplan–Meier estimator of the survival function, also known as the product limit estimator, is a non-parametric statistic used to estimate the survival function from lifetime data. In medical research, it is often used to measure the fraction of patients living for a certain amount of time after treatment. More general alternatives to this are the Cox proportional hazards (PH) model and tree based models, which can also help identify important covariates for estimating the survival function.
The Cox PH regression model is a linear model. It is similar to linear regression and logistic regression. Specifically, these methods assume that a single line, curve, plane, or surface is sufficient to separate groups (alive, dead) or to estimate a quantitative response (survival time).
In some cases alternative partitions give more accurate classification or quantitative estimates.

One set of alternative methods are tree-structured survival models, including survival random forests. Tree-structured survival models may give more accurate predictions than Cox models. An alternative to building a single survival tree is to build many survival trees, where each tree is constructed using a sample of the data, and average the trees to predict survival. This is the method underlying the survival random forest models. The prediction errors are estimated by bootstrap re-sampling [9]. Random Survival Forest (RSF) are an extension of Random Forest to analyze right censored, time to event data. A forest of survival trees is grown using a log-rank splitting rule to select the optimal candidate variables. Survival estimate for each observation are constructed with a Kaplan–Meier (KM) estimator within each terminal node, at each event time. Random Survival Forests adaptively discover nonlinear effects and interactions and are fully nonparametric. Averaging over many trees enables RSF to approximate complex survival functions, including non-proportional hazards, while maintaining low prediction error. It has been shown [9] that RSF is uniformly consistent and that survival forests have a uniform approximating property in finite-sample settings, a property not possessed by individual survival trees.

Previous work to identify heterogeneity in survival and form sub-groups includes building generative models (such as Gaussian mixtures) and clustering to form homogenous groups with units which have similar survival curves [7]. Stratified Cox models and frailty functions have also been proposed [8]. Also, the above mentioned tree/random forest based models have been used earlier with good results and favorable theoretical properties [9]. However, these methods do not help identify covariates which are causally important in determining the survival difference between two or more treatments, neither at the population not at a sub-group or individual unit level (heterogeneous treatment effects), i.e. they do not build upon a causal inference framework to estimate survival. In contrast, our proposed algorithm is at the intersection of causal inference and survival estimation using machine learning methods, specifically random forest ensembles, and is described in section 4.

## 3. Details of the dataset

The dataset used for this analysis was collected from multiple cancer treatment centers across different cities in India, and was focused on breast cancer patients (1806 total). We look at two treatments: treatment T0: Surgery plus chemotherapy plus radiation (1286 patients), versus treatment T1: Surgery plus chemo only (520 patients). Another group of patients (262) tagged as terminally ill underwent treatment T2: palliative chemo. only, and is only included for preliminary survival pots, and excluded from further analyses.

There are 110 patient covariates/features, including features related to demographics (age, gender, etc.), cancer location, tumor grade, cancer stage, biomarker (ER,PR,HER2) status, metastasis status, menopausal status, type of surgery/chemo/radiation, various drugs administered and dosages, hormonal therapy status, and so on. For each patient, the disease free survival days (DFS) and death event binary indicator was recorded. Our aim is to identify covariates which pick treatment effect heterogeneity, and form groups based on that criterion. Then, for each leaf/split group, we identify the combination of covariates leading to that. Then, we proceed to predict survival under both treatments for each patient, and compute the difference. Details are outlined below.

## 4. Treatment effects for survival: Proposed algorithm details & results

**Preliminary analysis: Patient segmentation based on survival curve estimation and prediction**

We analyze patient sub-populations who underwent different treatments, and also estimate the survival at an individual patient level using all the patient features. This can be very useful for new patients, for whom we can predict their survival curves (for each treatment plan) using this model before they are assigned a particular treatment.

We can also segment patient sub-groups in each treatment, based on their estimated survival, and look at features which are important in forming these clusters with different treatment effect (T1-T0).

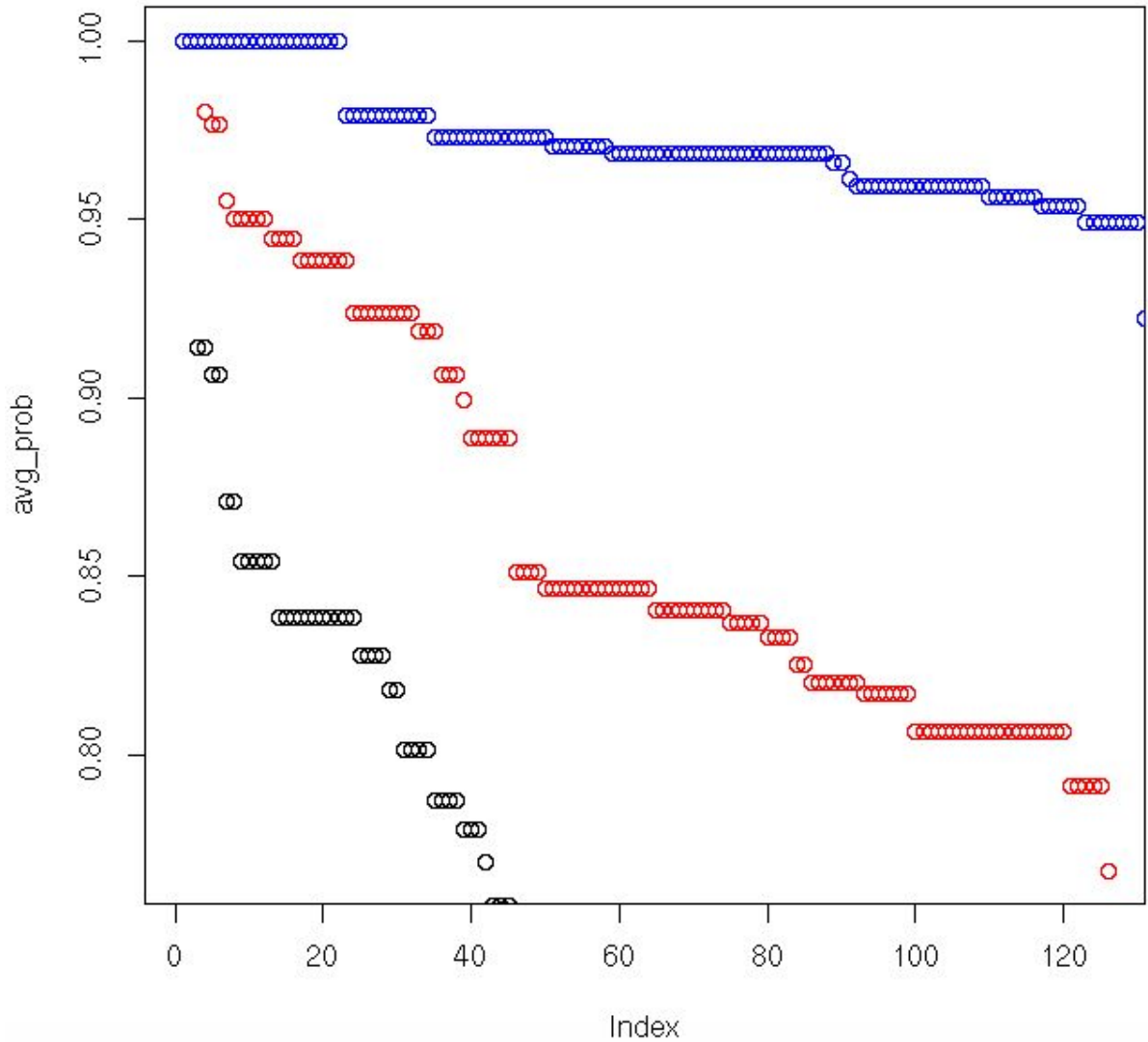

This plot shows the average patient survival for different treatment plans.
Blue: Treatment 1 (best survival)
Red:Treatment 0 (medium survival)
Black:Treatment 2 (lowest survival): not used in further analysis.
There is a clear difference the average survival curves across treatments.

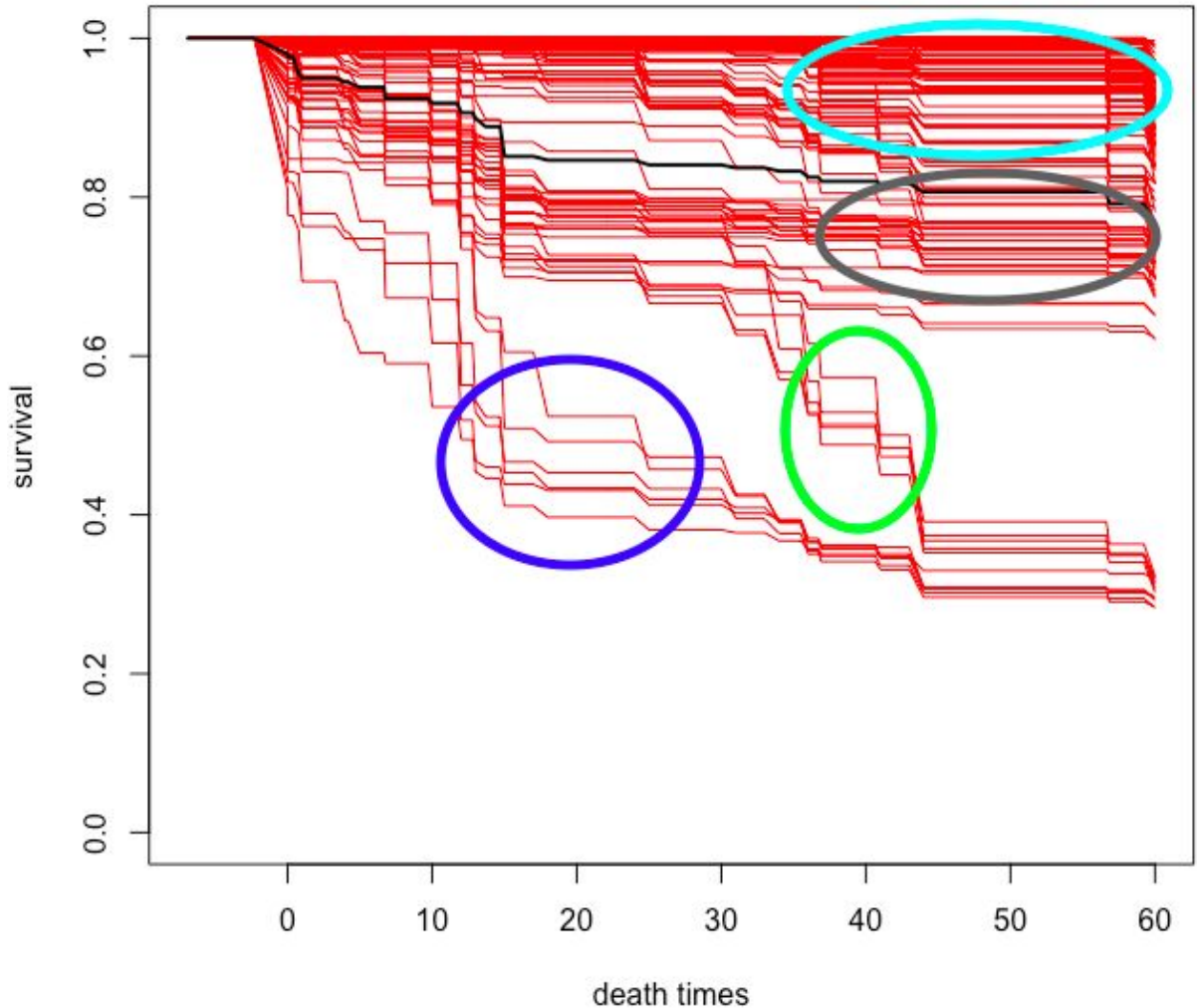

The above survival plot is for **Treatment T1 only, per patient.**

As can be seen above, even for a single treatment plan, there is reasonable heterogeneity in survival, determined by various patient features and other factors. Each red plot is for an individual patient (predicted survival), black is the average.

For T1, in the above plot, we can see 4 patient subgroups based on survival, shown by differently colored circles manually overlaid for visualization (the clusters of red survival curves which are close together).

As preliminary analysis, we fit a **survival random forest** to each treatment type.

The model has 92% accuracy on average in predicting survival curves, using the survival forests model.

For treatment T1, one set of patient features found which help differentiate between these sub-groups are **metastasis status, metastasis sites, pathology stage, menopausal status, PR status, chemo. Type and radiation dose.** Thus, survival forests can predict survival curves well for individual patients given a treatment type. However, this analysis does not tell us anything about the features which are important for differential survival between two treatments for the same patient as well as patient sub-groups. We propose an algorithm to do exactly that: **In step 1: we will perform a causalTree analysis** to identify features for subgroups which result in a difference in disease free days (DFS outcome) for one treatment versus another (causal inference modeling), and then, **in step 2, fit survival forests per sub-group (leaf), per each treatment** to estimate the per patient and patient subgroup gain or loss difference in survival between the two treatments, conditioned on the features identified in the first step, causal inference.

## **Proposed algorithm Step 1: Causal inference using random forests: Application to survival curves**

Computing the median survival values for Treatment 0: S+CT (34 days) versus Treatment 1: S+CT+RT (43 days), with the median difference = (43-34) = 9 days, across the patient sub-population is not very useful to determine sub-groups and heterogeneity. In this section, we ask the following question:
Can we get a better fine grained difference of survival values between treatments T0 and T1, and identify the patient sub-groups as well as their important features which lead to this difference.
Using the causal inference framework (Rubin causal model), we have a missing data problem here, i.e. we know the outcome (=survival value) for only one treatment for each patient. This is well suited for a causalTree/causalForest model to estimate individual treatment effects (Survival_Treatment1- Survival_Treatment0) for individual patients.

Thus, in step 1 we fit a causalTree [5] to our dataset described above. Below one such tree is plotted, and the top splitting patient variables are important, with the values in the boxes giving the survival days difference between treatments for each patient subgroup (leaf).

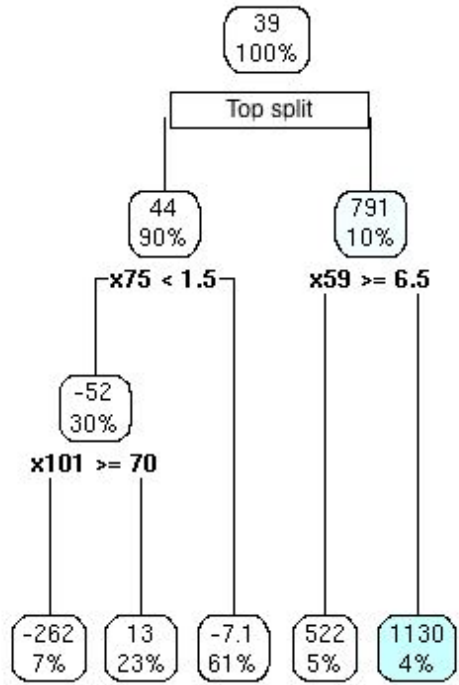

Figure: CausalTree for 2 treatments

Note that some of the survival difference values in the boxes for some sub-groups are quite different from the median difference value between T1 and T0.

Some **causally** important variables (in the above tree) are: **Chemo. chemical type used, radiation type, surgery type, and ER biomarker status.**

### Proposed algorithm step 2: Survival forests per sub-group/leaf

For each selected leaf, we fit 2 survival random forests: one for each treatment, for all patients in that leaf. To illustrate this, for one such sub-group (tree leaf segment sub-population), specifically, for the leftmost bottom split in the causalTree figure (The path being Top_split→ x_75<1.5 → x101>=70).

For all the patients in this subgroup/leaf, we fit 2 survival random forests: One forest for treatment T1 and one forest for treatment T0, both using the subset of features identified by the causalTree, and predict the survival curves for a test subset in that leaf. The 2 sub-plots below show the patient level survival curves for the two treatments, and the third sub-plot shows the overlay, and the fourth/last sub-plot shows the estimated/predicted difference in the survival curve for one of the chosen patients (the arrows show the difference between the red-blue dotted curves). Thus, after step 2, we can identify the survival difference curve between treatments T0 and T1 for the leaf, as well as patient features/variables causing this difference.

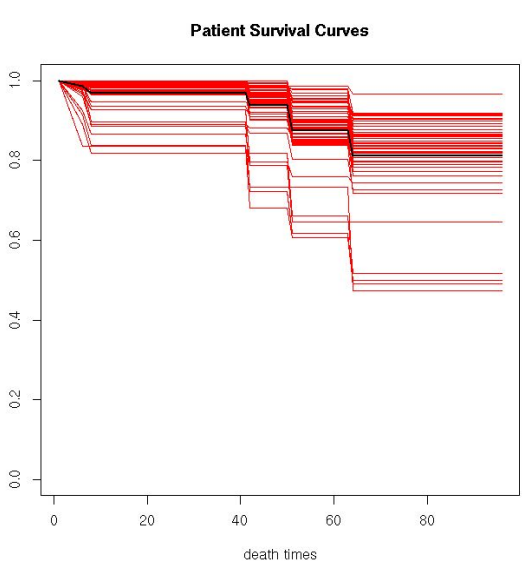
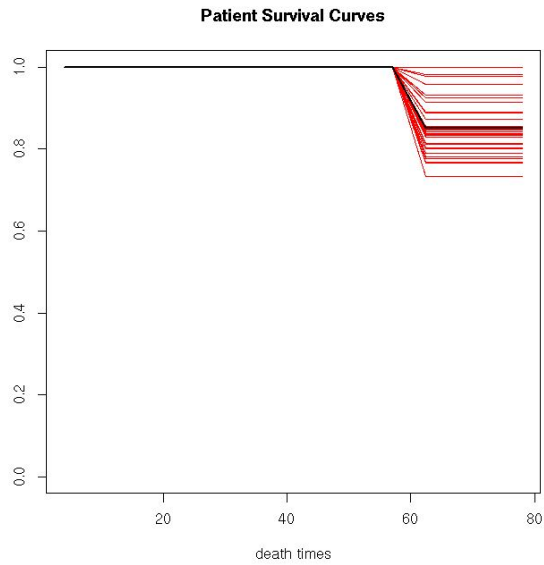

Treatment 0: survival for sub-group (left)     Treatment 1: survival for sub-group (right)

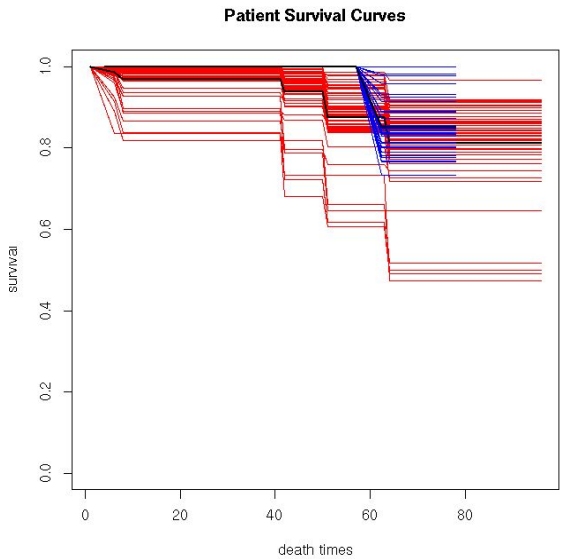
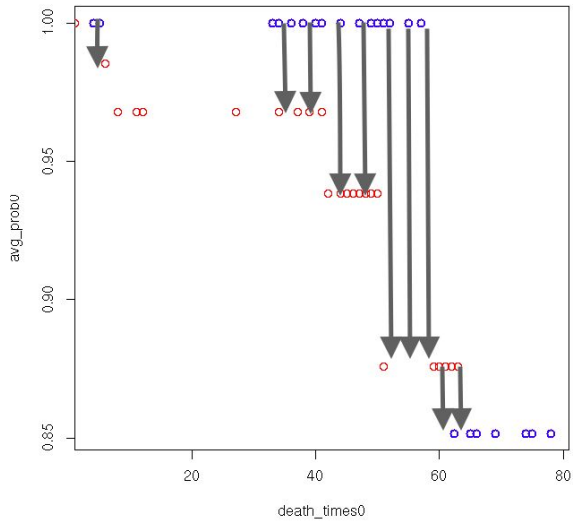

Survival curves overlaid for both treatments     Differential survival curve at patient level

Figure: Top left: Survival curve for T0 in chosen leaf, Top right: survival curve for T1 in chosen leaf (note that both are more homogenous than the survival curve for full dataset),
Bottom left: Overlay of two treatments for a patient sub-group (left)
Bottom right: Predicted difference in survival for a given patient in the leaf, for the two different treatments (right): arrows show the area under the difference curve.

## 5. Discussion and conclusion

Recently, there have been several efforts to leverage machine learning techniques for causal inference problems, including estimating heterogeneous treatment effects [5], propensity score modeling as well as neighbor matching [1] for individual treatment effects. Our aim is to contribute to this by extending causal inference and heterogeneous treatment effect estimation (at individual and sub-group level) where the outcome is survival. We have outlined the 2 step algorithm. In step 1, we use a causalTree/causalForest to partition the data into different sub-groups/leaves based on the treatment effect differences for survival days. Then, in step 2, we build two survival random forests at individual leaves (one for each treatment), and predict survival under both treatments for each patient, as well as the differential survival curve. Code for the algorithm will be made available shortly on Github at this location: https://github.com/vikas84bf

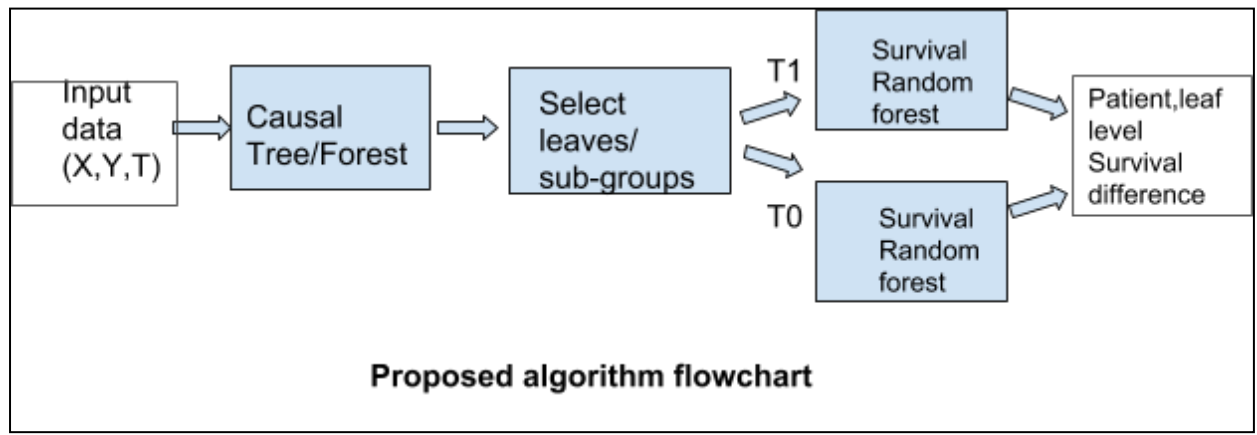

Figure: A flowchart illustrating the proposed algorithm

## Acknowledgement

We would like to thank Prof. Susan Athey and Prof. Guido Imbens at the Stanford University GSB for several illuminating discussions about causal inference, treatment effects and econometrics.